\documentstyle[aps,epsf]{revtex}  

\newcommand{\plb}[2]{{\em Phys. Lett.}              {\bf #1B}, #2 }
\newcommand{\npb}[2]{{\em Nucl. Phys.}              {\bf B#1}, #2 }

\newcommand{\qrd}[2]{{\em Phys. Rev.}               {\bf D#1}, #2 }
\newcommand{\qrl}[2]{{\em Phys. Rev. Lett.}         {\bf  #1}, #2 }
\newcommand{\zpc}[2]{{\em Z. Phys.}                 {\bf C#1}, #2 }

\newcommand{\epj}[2]{{\em Eur. Phys. J.}            {\bf C#1}, #2 }
\newcommand{\con}[2]{                               {\bf  #1}, #2 }
\newcommand{\etal}{{\em et al.}}
\newcommand{\ibid}{{\em ibid.}}
\newcommand{\col}{Collaboration}
\newcommand{\be}{\begin{equation}}
\newcommand{\ee}{\end{equation}}
\newcommand{\ba}{\begin{array}}
\newcommand{\ea}{\end{array}}
\newcommand{\ms}{{\overline{\rm MS}}}
\newcommand{\lsim}{\buildrel < \over {_\sim}}
\newcommand{\gsim}{\buildrel > \over {_\sim}}

\begin{document}

\baselineskip 14pt

\title{Implications of Precision Electroweak Measurements for Physics Beyond 
       the SM\thanks{Talk presented at the Division of Particles and Fields 
       Conference (DPF 99), Los Angeles, CA, January 5--9, 1999.}}
\author{Jens Erler}
\address{University of Pennsylvania}
\maketitle

\begin{abstract}
We perform a global analysis of electroweak precision measurements to find 
constraints on physics beyond the Standard Model. In particular, we discuss 
oblique parameters, which are useful to constrain additional matter
fields, as well as extra $Z$ bosons, and supersymmetry. We also summarize 
the present information on the Higgs boson mass. 
\end{abstract}

\section{Introduction}
Using the top quark and $Z$ boson masses, $m_t$ and $M_Z$, 
the QED coupling, $\alpha$, and the Fermi constant, $G_F$, as input, other 
precision observables can be computed within the SM as functions of
the Higgs boson mass, $M_H$. For relatively low values of $M_H$, the agreement 
with the measurements is found to be excellent, establishing the SM at the 
one-loop level. I will briefly review the constraints on $M_H$ and the 
experimental situation before moving beyond the SM.

Besides the recent high precision measurements of the $W$ boson 
mass~\cite{Karlen98,Dorigo98}, $M_W$,
the most important input into precision tests of electroweak theory continues 
to come from the $Z$ factories LEP~1~\cite{Karlen98} and SLC~\cite{Baird98}. 
The vanguard of the physics program at LEP~1 with about 20 million recorded $Z$
events is the analysis of the $Z$ lineshape. Its parameters are $M_Z$, the 
total $Z$ width, $\Gamma_Z$, the hadronic peak cross section, 
$\sigma_{\rm had}$, and the ratios of hadronic to leptonic decay widths, 
$R_\ell = {\Gamma({\rm had})\over \Gamma(\ell^+\ell^-)}$, where $\ell = e$, 
$\mu$, or $\tau$. They are determined in a common fit with the leptonic 
forward-backward (FB) asymmetries, $A_{FB} (\ell) = {3\over 4} A_e A_\ell$. 
With $f$ denoting the fermion index, 
\be
  A_f = {2 v_f a_f\over v_f^2 + a_f^2}
\ee
is defined in terms of the vector 
($v_f = I_{3,f} - 2 Q_f \sin^2 \theta_f^{\rm eff}$) and axial-vector 
($a_f = I_{3,f}$) $Zf\bar{f}$ coupling; $Q_f$ and $I_{3,f}$ are the electric 
charge and third component of isospin, respectively, and 
$\sin^2 \theta_f^{\rm eff} \equiv \bar{s}^2_f$ is an effective mixing angle.

An average of about 73\% polarization of the electron beam at the SLC allows 
for a set of competitive and complementary measurements with a much smaller 
number of $Z$'s ($\gsim 500,000$). In particular, the left-right (LR) cross 
section asymmetry, $A_{LR} = A_e$, represents the most precise determination 
of the weak mixing angle by a single experiment (SLD)~\cite{Baird98}. 
Mixed FB-LR asymmetries, $A^{FB}_{LR} (f) = {3\over 4} A_f$, single out the 
final state coupling of the $Z$ boson. 

For several years there has been an experimental discrepancy at the $2 \sigma$ 
level between $A_\ell$ from LEP and the SLC. With the 1997/98 high statistics 
run at the SLC, and a revised value for the FB asymmetry of the $\tau$ 
polarization, ${\cal P}^{FB}_\tau$, the two determinations are now consistent 
with each other,
\be \ba{l}
\label{aell}
  A_\ell ({\rm LEP}) = 0.1470 \pm 0.0027, \\
  A_\ell ({\rm SLD}) = 0.1503 \pm 0.0023.
\ea \ee
The LEP value is from $A_{FB}(\ell)$, ${\cal P}_\tau$, and ${\cal P}^{FB}_\tau$,
while the SLD value is from $A_{LR}$ and $A^{FB}_{LR} (\ell)$. The data is 
consistent with lepton universality, which is assumed here. There remains,
however, a $2.5 \sigma$ discrepancy between the two most precise 
determinations of $\bar{s}^2_\ell$, namely $A_{LR}$ and $A_{FB} (b)$ 
(assuming no new physics in $A_b$). 

Of particular interest are the results on the heavy flavor 
sector~\cite{Karlen98} including 
$R_q = {\Gamma (q\bar{q}) \over \Gamma ({\rm had})}$, $A_{FB} (q)$, and 
$A^{FB}_{LR} (q)$, with $q = b$ or $c$. There is a theoretical 
prejudice that the third family is the one which is most likely affected by 
new physics. Interestingly, the heavy flavor sector has always shown the 
largest deviations from the SM. E.g., $R_b$ deviated at times by almost 
$4 \sigma$. Now, however, $R_b$ is in good agreement with the SM, and thus 
puts strong constraints on many types of new physics. At present, there is some
discrepancy in $A^{FB}_{LR} (b) = {3\over 4} A_b$, and 
$A_{FB} (b) = {3\over 4} A_e A_b$, both at the $2 \sigma$ level. Using 
the average of Eqs.~(\ref{aell}), $A_\ell = 0.1489 \pm 0.0018$, both can be 
interpreted as measurements of $A_b$. From $A_{FB} (b)$ one would obtain 
$A_b = 0.887 \pm 0.022$, and the combination with 
$A^{FB}_{LR} (b) = {3\over 4} (0.867 \pm 0.035)$ would yield 
$A_b = 0.881 \pm 0.019$, which is almost $3 \sigma$ below the SM prediction. 
Alternatively, one could use $A_\ell ({\rm LEP})$ above (which is closer to the
SM prediction) to determine $A_b ({\rm LEP}) = 0.898 \pm 0.025$, and 
$A_b = 0.888 \pm 0.020$ after combination with $A^{FB}_{LR} (b)$, i.e., still 
a $2.3 \sigma$ discrepancy. An explanation of the 5--6\% deviation in $A_b$
in terms of new physics in loops, would need a 25--30\% radiative correction to 
$\hat\kappa_b$, defined by 
$\bar{s}^2_b \equiv \hat\kappa_b\sin^2\hat\theta_\ms (M_Z) \equiv \hat{s}^2_Z$.
Only a new type of physics which couples at the tree level 
preferentially to the third generation~\cite{Erler95}, and which does not 
contradict $R_b$ (including the off-peak measurements by 
DELPHI~\cite{Abreu96}), can conceivably account for a low $A_b$. 
Given this and that none of the observables deviates by $2 \sigma$ or more, 
we can presently conclude that there is no compelling evidence for new physics 
in the precision observables, some of which are listed in Table~\ref{zpole}. 
Very good agreement with the SM is observed. Only $A_{LR}$ and the two 
measurements sensitive to $A_b$ discussed above, show some 
deviation, but even those are below $2\sigma$.

\begin{table}[h]
\caption{Principal precision observables from CERN, FNAL, SLAC, and elsewhere.
Shown are the experimental results, the SM predictions, and the pulls. 
The SM errors are from the 
uncertainties in $M_Z$, $\ln M_H$, $m_t$, $\alpha (M_Z)$, and $\alpha_s$. 
They have been treated as Gaussian and their correlations have been taken into 
account. $\bar{s}_\ell^2 (Q_{FB} (q))$ is the weak mixing angle from the 
hadronic charge asymmetry; $R^-$ and $R^\nu$ are cross section ratios from 
deep inelastic $\nu$-hadron scattering; $g_{V,A}^{\nu e}$ are effective 
four-Fermi coefficients in $\nu$-e scattering; and the $Q_W$ are the weak 
charges from parity violation measurements in atoms. The uncertainty in the 
$b\rightarrow s\gamma$ observable includes theoretical errors from the physics
model, the finite photon energy cut-off, and from uncalculated higher order 
effects. There are other precision observables which are not shown but 
included in the fits.}
\label{zpole}
\begin{tabular}{lcccr}
Quantity & Group(s) & Value & Standard Model & pull \\ 
\hline
$M_Z$       \hspace{0pt} [GeV]&     LEP     &$ 91.1867 \pm 0.0021 $&$ 91.1865 \pm 0.0021 $&$ 0.1$ \\
$\Gamma_Z$  \hspace{3pt} [GeV]&     LEP     &$  2.4939 \pm 0.0024 $&$  2.4957 \pm 0.0017 $&$-0.8$ \\
$\sigma_{\rm had}$       [nb] &     LEP     &$ 41.491  \pm 0.058  $&$ 41.473  \pm 0.015  $&$ 0.3$ \\
$R_e$                         &     LEP     &$ 20.783  \pm 0.052  $&$ 20.748  \pm 0.019  $&$ 0.7$ \\
$R_\mu$                       &     LEP     &$ 20.789  \pm 0.034  $&$ 20.749  \pm 0.019  $&$ 1.2$ \\
$R_\tau$                      &     LEP     &$ 20.764  \pm 0.045  $&$ 20.794  \pm 0.019  $&$-0.7$ \\
$A_{FB} (e)$                  &     LEP     &$  0.0153 \pm 0.0025 $&$  0.0161 \pm 0.0003 $&$-0.3$ \\
$A_{FB} (\mu)$                &     LEP     &$  0.0164 \pm 0.0013 $&$                    $&$ 0.2$ \\
$A_{FB} (\tau)$               &     LEP     &$  0.0183 \pm 0.0017 $&$                    $&$ 1.3$ \\
\hline
$R_b$                         &  LEP + SLD  &$  0.21656\pm 0.00074$&$  0.2158 \pm 0.0002 $&$ 1.0$ \\
$R_c$                         &  LEP + SLD  &$  0.1735 \pm 0.0044 $&$  0.1723 \pm 0.0001 $&$ 0.3$ \\
$A_{FB} (b)$                  &     LEP     &$  0.0990 \pm 0.0021 $&$  0.1028 \pm 0.0010 $&$-1.8$ \\
$A_{FB} (c)$                  &     LEP     &$  0.0709 \pm 0.0044 $&$  0.0734 \pm 0.0008 $&$-0.6$ \\
$A_b$                         &     SLD     &$  0.867  \pm 0.035  $&$  0.9347 \pm 0.0001 $&$-1.9$ \\
$A_c$                         &     SLD     &$  0.647  \pm 0.040  $&$  0.6676 \pm 0.0006 $&$-0.5$ \\
\hline
$A_{LR} + A_\ell$             &     SLD     &$  0.1503 \pm 0.0023 $&$  0.1466 \pm 0.0015 $&$ 1.6$ \\
${\cal P}_\tau: A_e+A_\tau$   &     LEP     &$  0.1452 \pm 0.0034 $&$                    $&$-0.4$ \\
$\bar{s}_\ell^2 (Q_{FB})$     &     LEP     &$  0.2321 \pm 0.0010 $&$  0.2316 \pm 0.0002 $&$ 0.5$ \\
\hline
$m_t$      \hspace{6pt} [GeV]&  Tevatron   &$173.8    \pm 5.0    $&$171.4    \pm 4.8    $&$ 0.5$ \\
$M_W$      \hspace{0pt}  [GeV]&     all     &$ 80.388  \pm 0.063  $&$ 80.362  \pm 0.023  $&$ 0.4$ \\
\hline
$R^-$          &     NuTeV      &$   0.2277 \pm 0.0021 \pm 0.0007 $&$   0.2297 \pm 0.0003 $&$-0.9$\\
$R^\nu$        &     CCFR       &$   0.5820 \pm 0.0027 \pm 0.0031 $&$   0.5827 \pm 0.0005 $&$-0.2$\\
$R^\nu$        &     CDHS       &$   0.3096 \pm 0.0033 \pm 0.0028 $&$   0.3089 \pm 0.0003 $&$ 0.2$\\
$R^\nu$        &     CHARM      &$   0.3021 \pm 0.0031 \pm 0.0026 $&$                     $&$-1.7$\\
\hline
$g_V^{\nu e}$  &      all       &$  -0.041  \pm 0.015             $&$  -0.0395 \pm 0.0004 $&$-0.1$\\
$g_A^{\nu e}$  &      all       &$  -0.507  \pm 0.014             $&$  -0.5063 \pm 0.0002 $&$-0.1$\\
\hline
$Q_W({\rm Cs})$&     Boulder    &$ -72.41   \pm 0.25\pm 0.80      $&$ -73.10   \pm 0.04   $&$ 0.8$\\
$Q_W({\rm Tl})$&Oxford + Seattle&$-114.8    \pm 1.2 \pm 3.4       $&$-116.7    \pm 0.1    $&$ 0.5$\\
\hline
${\Gamma (b\rightarrow s\gamma)\over \Gamma (b\rightarrow c e\nu)}$& CLEO 
           &$ 3.26^{+0.75}_{-0.68} \times 10^{-3} $&$ 3.14^{+0.19}_{-0.18} 
           \times 10^{-3} $&$ 0.1$\\
\end{tabular}
\end{table}

The data show a strong preference for a low $M_H \sim {\cal O} (M_Z)$. 
Unlike in previous analyses, the central value of the global fit 
to all precision data, including $m_t$ and excluding further 
constraints from direct searches,
\be
\label{mh_fit}
   M_H = 107^{+67}_{-45} \mbox{ GeV},
\ee
is now above the direct lower limit, $M_H > 90 \mbox{ GeV [95\% CL]}$,
from searches at LEP 2~\cite{McNamara98}. It coincides with the
$5 \sigma$ discovery limit from LEP 2 running at 200~GeV center of mass energy 
with 200 pb$^{-1}$ integrated luminosity per experiment~\cite{McNamara98}. 
The 90\% central confidence interval from precision data only is given by
$39 \mbox{ GeV} < M_H < 226 \mbox{ GeV}$. The fit result~(\ref{mh_fit}) is 
consistent with the predictions for the lightest neutral Higgs 
boson~\cite{Carena96}, $m_{h^0} \lsim 130$ [150]~GeV,
within the Minimal Supersymmetric Standard Model (MSSM) [and its extensions].

For the determination of the proper $M_H$ upper limits, we scan equidistantly 
over $\ln M_H$, combining the likelihood $\chi^2$ function from the 
precision data with the exclusion curve (interpreted as a prior probability 
distribution function) from LEP 2~\cite{McNamara98}. This curve is from Higgs 
searches at center of mass energies up to 183 GeV. We find the 90 (95, 99)\% 
confidence upper limits,
\be
\label{mh_limits}
  M_H < 220 \mbox{ (255, 335) GeV}.
\ee
Notice, that the LEP 2 exclusion curve increases the 95\% upper limit by
almost 30 GeV. The upper limits (\ref{mh_limits}) are rather insensitive 
to the $\alpha(M_Z)$ used in the fits. This is due to 
compensating effects from the larger central value of $\alpha (M_Z)$ 
(corresponding to lower extracted Higgs masses) and the larger error bars 
in the data driven approach~\cite{Eidelman95} as compared to evaluations 
relying more strongly on perturbative QCD~\cite{Zeppenfeld95}. 
While the limits are therefore robust 
within the SM, it should be cautioned that the results on $M_H$ are strongly 
correlated with certain new physics parameters, as discussed in 
Section~\ref{oblique}.

The accurate agreement of theory and experiment allows severe constraints on 
possible TeV scale physics, such as unification or compositeness. For example,
the ideas of technicolor and non-supersymmetric Grand Unified Theories (GUTs)
are strongly disfavored. On the other hand, supersymmetric unification, as 
generically predicted by heterotic string theory, is supported by the observed 
approximate gauge coupling unification at an energy slightly below the Planck 
scale, and by the decoupling of supersymmetric particles from the precision 
observables. As I will discuss in the following Sections, those types of
new physics which tend to decouple from the SM are favored, while 
non-decoupling new physics generally conflicts with the data. 

\section{Oblique parameters: bounds on extra matter}
\label{oblique}
The data is precise enough to constrain additional parameters describing 
physics beyond the SM. Of particular interest is the $\rho_0$-parameter, 
which is a measure of the neutral to charged current interaction strength
and defined by
\be
  \rho_0 = {M_W^2 \over M_Z^2 \hat{c}^2_Z \hat{\rho} (m_t,M_H)}.
\ee
The SM contributions are absorbed in $\hat{\rho}$. Examples for sources of 
$\rho_0 \neq 1$ include {\it non-degenerate\/} extra fermion or boson 
doublets, and non-standard Higgs representations. 

In a fit to all data with $\rho_0$ as an extra fit parameter, there is a strong
(73\%) correlation\footnote{$\rho_0$ is also strongly anticorrelated with the
strong coupling $\alpha_s$ ($-53\%$) and $m_t$ ($-46\%$).} 
with $M_H$. As a result, upper limits on $M_H$ are weaker when $\rho_0$ is 
allowed. Indeed, $\chi^2 (M_H)$ is very shallow with 
$\Delta \chi^2 = \chi^2 (1 \mbox{ TeV}) - \chi^2 (M_Z) = 4.5$, and its minimum
is at $M_H = 46$~GeV, which is already excluded. For comparison, within the SM 
a 1~TeV Higgs boson is excluded at the 5~$\sigma$ level. We obtain,
\be
\ba{lcr}
\label{rhofit}
         \rho_0 &=& 0.9996^{+0.0009}_{-0.0006}, \vspace{3pt} \\
            m_t &=& 172.9   \pm 4.8 \mbox{ GeV}, \vspace{3pt} \\
 \alpha_s (M_Z) &=& 0.1212  \pm 0.0031,
\ea
\ee
in excellent agreement with the SM ($\rho_0 = 1$). The central values are for 
$M_H = M_Z$, and the uncertainties are $1 \sigma$ errors and include the range,
$M_Z \leq M_H \leq 167$~GeV, in which the minimum $\chi^2$ varies within one 
unit. Note, that the uncertainties for $\ln M_H$ and $\rho_0$ are non-Gaussian:
at the $2 \sigma$ level ($\Delta \chi^2 \leq 4$), Higgs boson masses up to 
800~GeV are allowed, and we find
\be
   \rho_0 = 0.9996^{+0.0031}_{-0.0013} \mbox{ ($2 \sigma$)}.
\ee
This implies strong constraints on the mass splittings of extra fermion and 
boson doublets~\cite{Veltman77}, 
\be
  \Delta m^2 = m_1^2 + m_2^2 - \frac{4 m_1^2 m_2^2}{m_1^2 - m_2^2} 
               \ln {m_1\over m_2} \geq (m_1 - m_2)^2,
\ee
namely, at the $1\sigma$ and $2\sigma$ levels, respectively, 
($C_i$ is the color factor)
\be
\label{splittings}
   \sum\limits_i {C_i\over 3} \Delta m^2_i < \mbox{ (38 GeV)}^2 
   \mbox{ and (93 GeV)}^2.
\ee
Due to the restricted Higgs mass range in the presence of supersymmetry
(SUSY), stronger $2 \sigma$ constraints result here,
\be
   \rho_0 \mbox{ (MSSM) } = 0.9996^{+0.0017}_{-0.0013} \mbox{ ($2 \sigma$)}.
\ee
The $2 \sigma$ constraint in~(\ref{splittings}) would therefore tighten from
$\mbox{(93 GeV)}^2$ to $\mbox{(64 GeV)}^2$. 

Constraints on heavy {\it degenerate chiral fermions\/} can be obtained 
through the $S$ parameter~\cite{Peskin90}, defined as a difference of $Z$ 
boson self-energies,
\be
  \frac{\hat\alpha (M_Z) }{4 \hat{s}_Z^2 \hat{c}_Z^2} S \equiv 
  \frac{\Pi^{\rm new}_{ZZ} (M_Z^2) - \Pi^{\rm new}_{ZZ} (0) }{M_Z^2}.
\ee
The superscripts indicate that $S$ includes new physics contributions only.
Likewise, $T = (1 - \rho_0^{-1})/\hat\alpha$ and the third oblique parameter,
$U$, also vanish in the SM. A fit to all data with $S$ allowed yields,
\be
\ba{lcr}
\label{Sfit}
              S &=& -0.20^{+0.24}_{-0.17}, \vspace{3pt} \\
            M_H &=& 390^{+690}_{-310} \mbox{ GeV}, \vspace{3pt} \\
            m_t &=& 172.9   \pm 4.8 \mbox{ GeV}, \vspace{3pt} \\ 
       \alpha_s &=& 0.1221  \pm 0.0035.
\ea
\ee
It is seen, that in the presence of $S$ constraints on $M_H$ virtually 
disappear. In fact, $S$ and $M_H$ are almost perfectly anticorrelated 
($-92\%$). By requiring $M_Z \leq M_H \leq 1$~TeV, we find at the $3\sigma$ level,
\be
   S = -0.20^{+0.40}_{-0.33} \mbox{ ($3 \sigma$)}.
\ee
A heavy degenerate ordinary or mirror family contributes $2/3\pi$ to $S$.
A degenerate fourth generation is therefore excluded at the 99.8\% CL
on the basis of the $S$ parameter alone. Due to the correlation with $T$, 
the fit becomes slightly better in the presence of a non-degeneracy of 
the new doublets. A non-vanishing $T = 0.15 \pm 0.08$ is favored, but
even in this case a fourth family is excluded at least at the 98.2\% CL. 
This is in agreement with a different constraint on the generation number,
using very different assumptions: allowing the invisible $Z$ width as a 
free parameter, yields the constraint, $N_\nu = 2.992 \pm 0.011$, on the 
number of light standard neutrino flavors.

A simultaneous fit to $S$, $T$, and $U$, can be performed only relative 
to a specified $M_H$. If one fixes $M_H = 600$~GeV, 
as is appropriate in QCD-like technicolor models, one finds
\be
\ba{rcr}
   S &=& -0.27 \pm 0.12, \\
   T &=&  0.00 \pm 0.15, \\
   U &=&  0.19 \pm 0.21.
\ea
\ee
Notice, that in such a fit the $S$ parameter is significantly smaller than 
zero. From this an isodoublet of technifermions, assuming $N_{TC} = 4$ 
technicolors, is excluded by almost 6 standard deviations, and a full 
technigeneration by more than $15\sigma$. However, the QCD-like models are 
excluded on other grounds, such as FCNC. These can be avoided in models of 
walking technicolor in which $S$ can also be smaller or even 
negative~\cite{Gates91}. 

\section{Extra $Z^\prime$ bosons}
Many GUTs and string models predict extra gauge symmetries and new exotic 
states. For example, $SO(10)$ GUT contains an extra $U(1)$ as can be seen
from its maximal subgroup, $SU(5) \times U(1)_\chi$. 
The $Z_\chi$ boson is also the unique solution to the conditions of
(i) no extra matter other than the right-handed neutrino,
(ii) absence of gauge and mixed gauge/gravitational anomalies, and
(iii) orthogonality to the hypercharge generator. 
Relaxing condition (iii) allows other solutions (including the $Z_{LR}$ 
appearing in left-right models with $SU(2)_L \times SU(2)_R \times U(1)$
gauge symmetry) which differ from the $Z_\chi$ boson by a shift proportional 
to the third component of the right-handed isospin generator~\cite{Erler99}. 
Equivalently, a non-vanishing kinetic mixing term~\cite{Babu98} can also 
parametrize these other solutions~\cite{Erler99}.

\begin{figure}[ht]
\centerline{\epsfxsize 2.35 truein 
\epsfbox{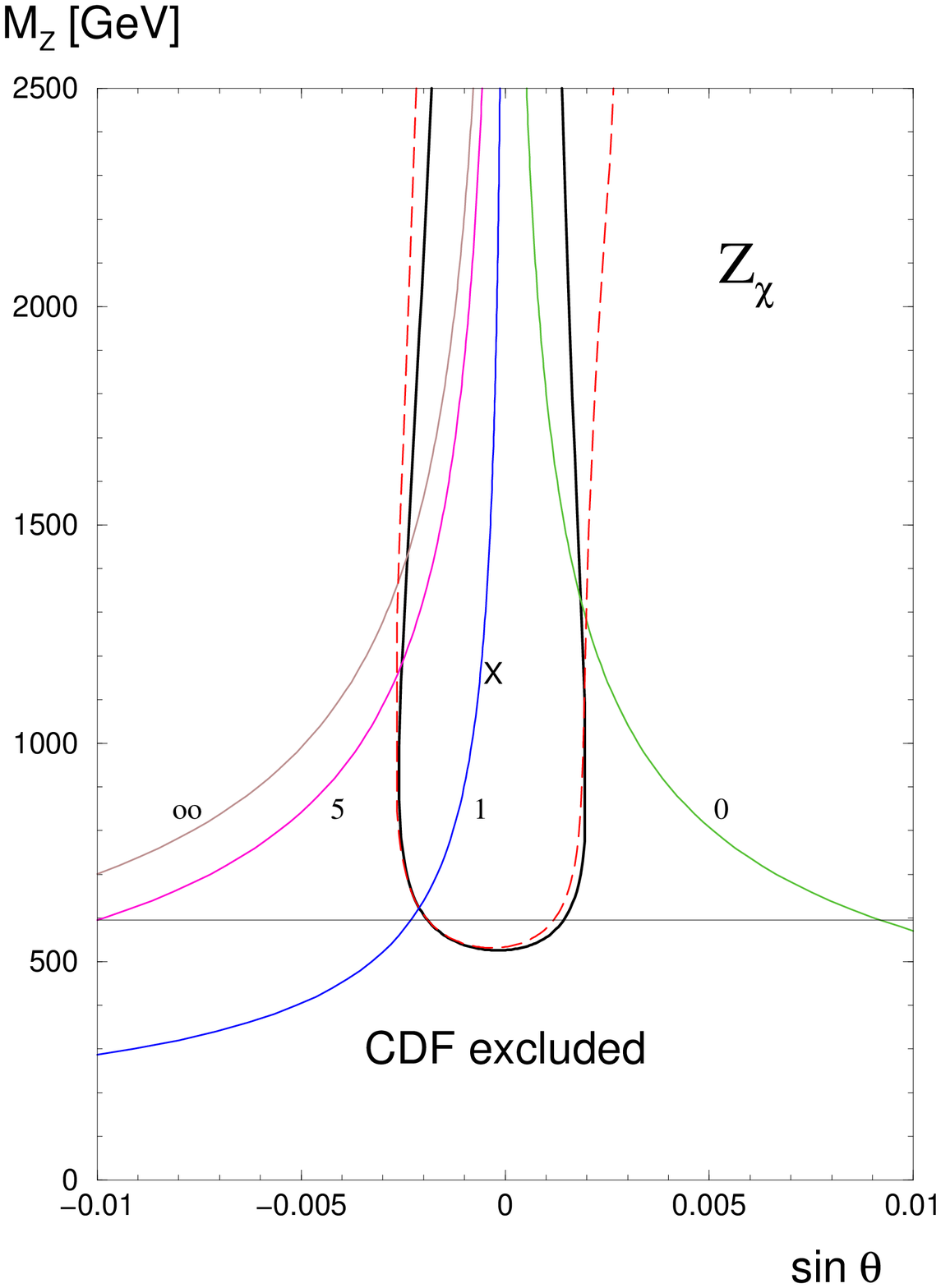}   
\epsfxsize 2.35 truein 
\epsfbox{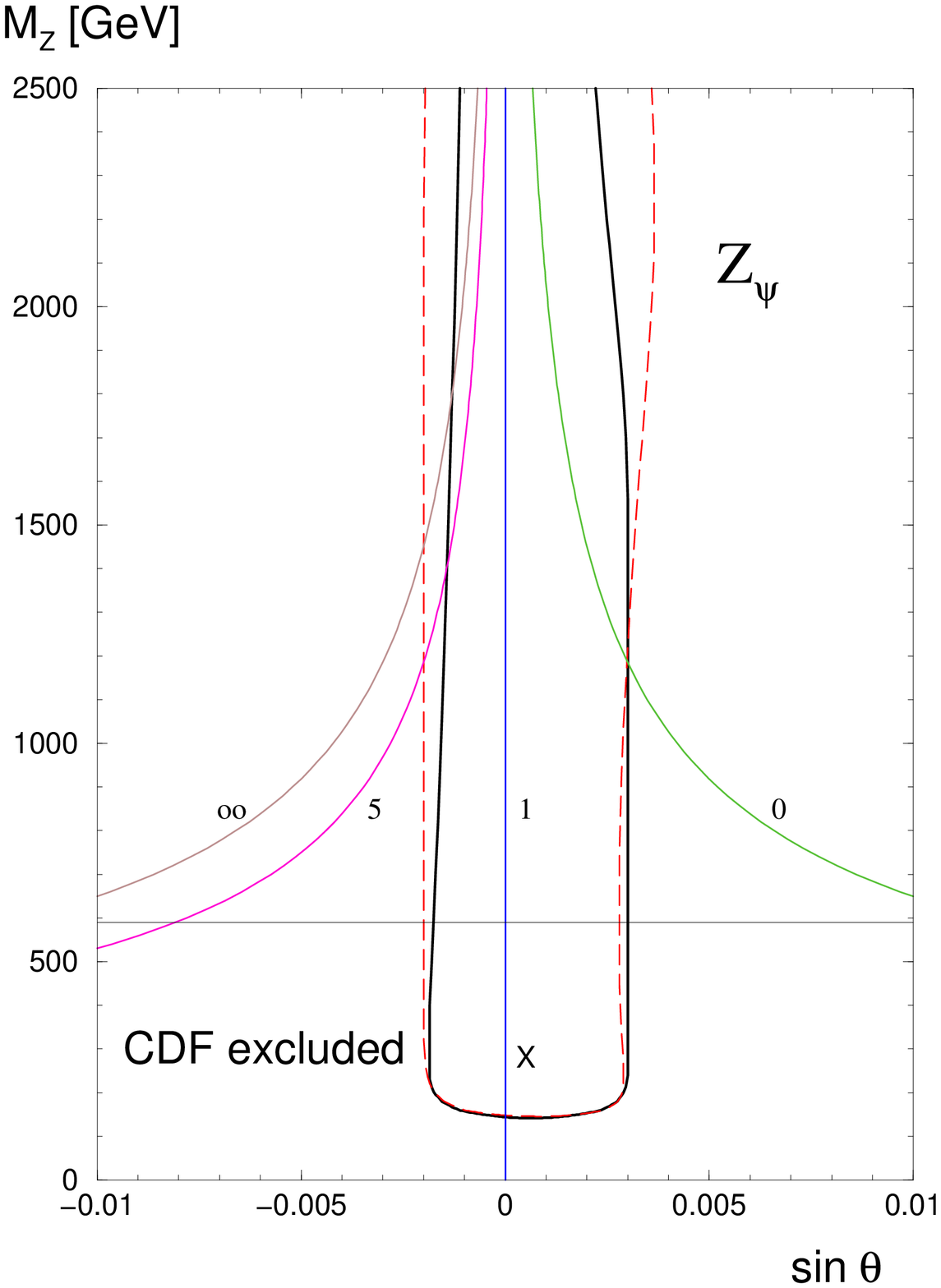}   
\epsfxsize 2.35 truein 
\epsfbox{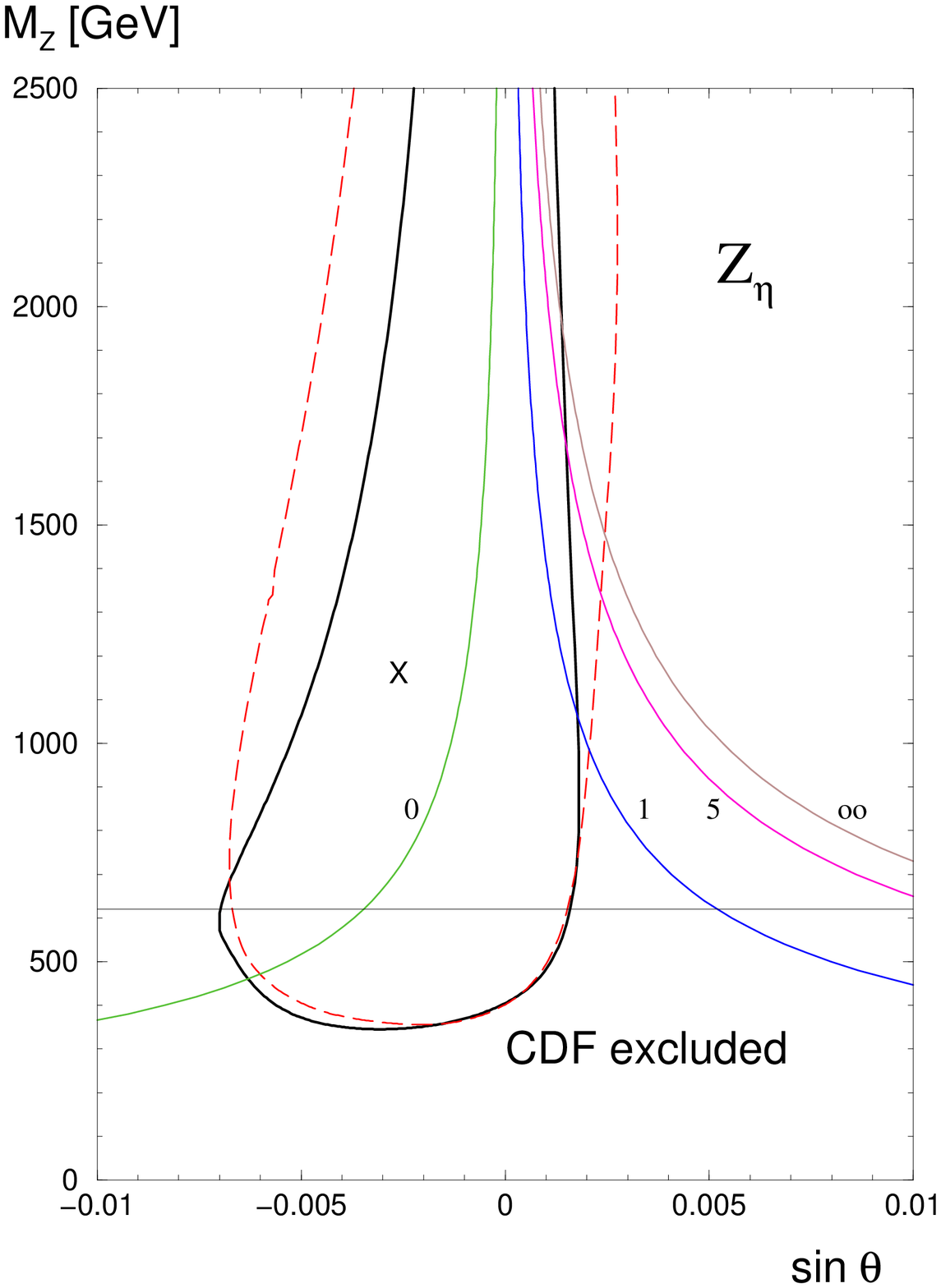}}   
\vskip .1 cm
\caption[]{
\label{zcontours}
\small 90\% CL contours for different $Z^\prime$ models. The solid 
contour lines use the constraint $\rho_0 = 1$ (the cross denotes the best fit 
for the $\rho_0 = 1$ case) while the long-dashed lines are for arbitrary 
Higgs sectors. Also shown are the additional constraints in minimal Higgs 
scenarios for several ratios of VEVs. The lower direct production limits from 
CDF~\cite{Abe97} are also shown.}
\end{figure}

Similarly, $E_6$ GUT contains the subgroup $SO(10) \times U(1)_\psi$, giving
rise to another $Z^\prime$. It possesses only axial-vector couplings to the 
ordinary fermions. As a consequence its mass, $M_{Z^\prime_\psi}$, is generally
less constrained (see Fig~\ref{zcontours}). 

The $Z_\eta$ boson is the linear combination 
$\sqrt{3/8}\, Z_\chi - \sqrt{5/8}\, Z_\psi$. It occurs in Calabi-Yau 
compactifications of the heterotic string if $E_6$ breaks directly to a rank~5 
subgroup~\cite{Witten85} via the Hosotani mechanism.

The potential $Z^\prime$ boson is in general a superposition of the SM $Z$ 
and the new boson associated with the extra $U(1)$. The mixing angle $\theta$
satisfies the relation~\cite{Langacker84},
\be
   \tan^2\theta = \frac{M_{Z_1^0}^2 - M_Z^2}{M_{Z^\prime}^2 - M_{Z_1^0}^2},
\ee
where $M_{Z_1^0}$ is the SM value for $M_Z$ in the absence of mixing. 
Note that $M_Z < M_{Z_1^0}$, and that the SM $Z$ couplings are changed by the
mixing. If the Higgs $U(1)^\prime$ quantum numbers are known, 
as well, there will be an extra constraint,
\be
  \theta = C {g_2\over g_1} {M_Z^2\over M_{Z^\prime}^2},
\ee
where $g_{1,2}$ are the $U(1)$ and $U(1)^\prime$ gauge couplings with
$g_2 = \sqrt{5\over 3}\, \sin \theta_W \sqrt{\lambda}\, g_1$.
$\lambda = 1$ (which we assume) if the GUT group breaks directly to
$SU(3) \times SU(2) \times U(1) \times U(1)^\prime$. $C$ is a function of 
vacuum expectation values (VEVs). For minimal Higgs sectors it can be
found in Table~III of reference~\cite{Langacker92}. Fig.~\ref{zcontours} 
shows allowed contours for $\rho_0$ free (see Section~\ref{oblique}), as well 
as $\rho_0 = 1$ (only Higgs doublets and singlets). Notice, that in the cases 
of minimal Higgs sectors the $Z^\prime$ mass limits are pushed into the TeV 
region. For more details and other examples see Ref.~\cite{Erler99}.

\section{Supersymmetry}

The good agreement between the SM predictions and the data favors
those types of new physics for which contributions decouple from the precision
observables. In particular, supersymmetric extensions of the SM with heavy
(decoupling) superpartners are in perfect agreement with observation. 
Other regions of parameter space, however, where some of the supersymmetric 
states are relatively light are strongly constraint by the data. 

In a recent analysis~\cite{Erler98} we systematically studied these 
constraints within the MSSM with various assumptions about the mediation 
of SUSY breaking (i.e.\ about the soft SUSY breaking terms). In a first
step, we identified the allowed region in parameter space taking into
account all direct search limits on superparticles, but ignoring the additional
information from the precision data. We then added the indirect constraints
arising from SUSY loop contributions. We found that a significant region of 
MSSM parameter space has to be excluded, and that the lower limits on 
superparticles and extra Higgs states strengthen. See the results in 
Fig.~\ref{susymasses} from an update of our analysis for this 
conference~\cite{Pierce99}. 

\section{Conclusions}
The precision data confirms the validity of the SM at the electroweak loop 
level, and there is presently no compelling evidence for deviations. A low 
Higgs mass is strongly favored by the data. While the precise range of $M_H$ 
is rather sensitive to $\alpha (M_Z)$, the upper limit is not. However, 
in the presence of non-standard contributions to the $S$ or $T$ parameters, 
no strong $M_H$ bounds can be found.

There are stringent constraints on parameters beyond the SM, such as
$S$, $T$, $U$, and others. This is a serious problem for models of dynamical 
symmetry breaking, compositeness, and the like, and excludes a fourth 
generation of quarks and leptons at the $3 \sigma$ level. Those 
constraints are, however, consistent with the MSSM, favoring its decoupling 
limit. Moreover, the low favored $M_H$ is in agreement with the expected mass 
range for the lightest neutral Higgs boson in the MSSM. Precision tests also
impose stringent limits on extra $Z^\prime$ bosons suggested in many GUT and 
string models. They limit their mixing with the ordinary $Z$, and put 
competitive lower limits on their masses, especially in concrete models 
in which the $U(1)^\prime$ charges of the Higgs sector are specified.

\begin{table}[ht]
\centerline{
\epsfxsize 6.5 truein 
\epsfbox{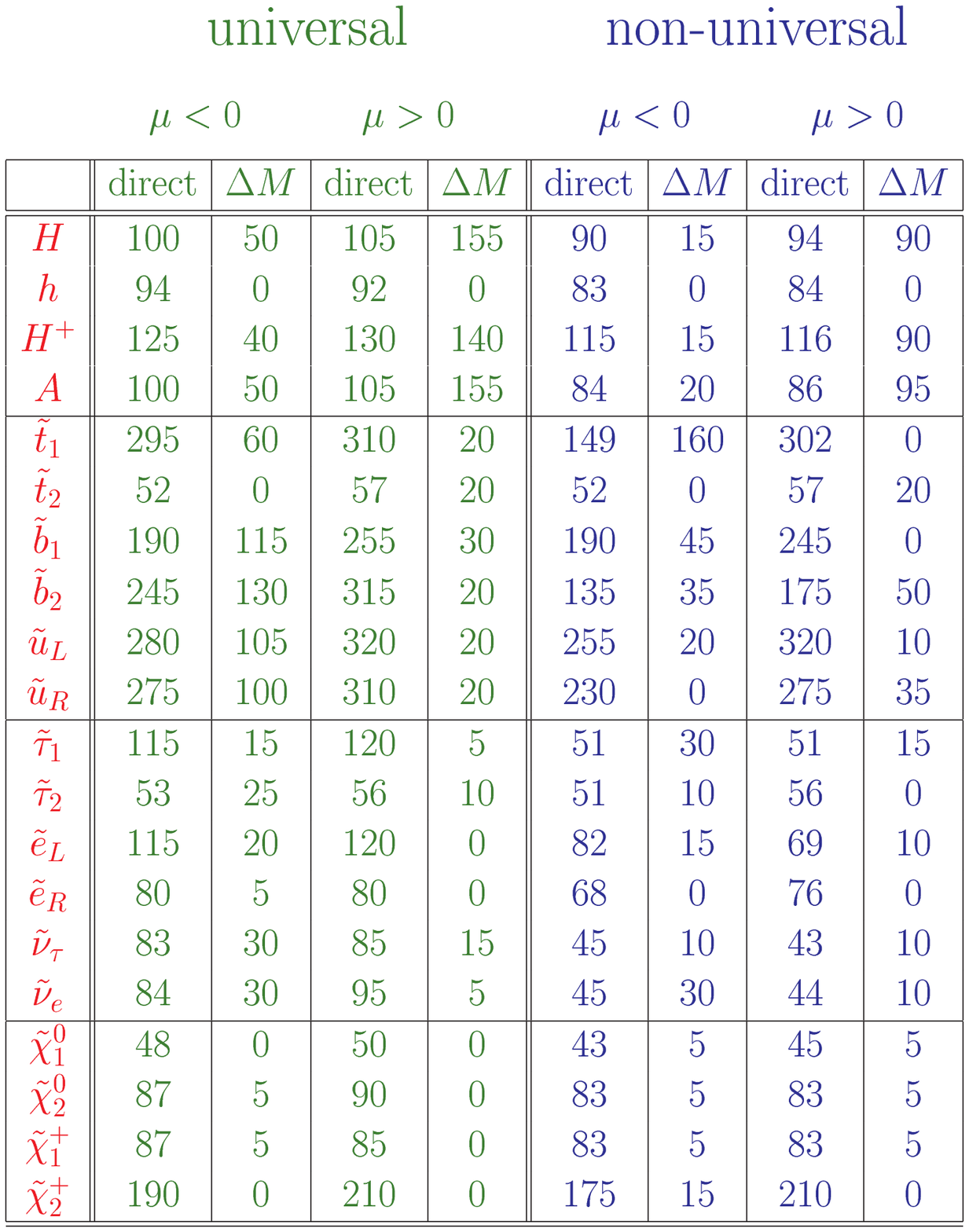}}   
\vspace{-2.cm}
\caption[]{
\label{susymasses}
\small Shifts ($\Delta M$) in the lower limits of superpartners and extra 
Higgs states. Considered are the two cases of universal and non-universal 
boundary conditions within the model of supergravity mediated SUSY breaking.
We separate the two cases of positive and negative sign of the 
supersymmetric bilinear Higgs ($\mu$) term.}
\end{table}

\section*{Acknowledgement}
It is a pleasure to thank Paul Langacker and Damien Pierce for collaboration.

\end{document}